\documentclass[authorversion,screen,nonacm]{acmart}
\setcopyright{none}

\acmConference[CHI 2022]{CHI 2022: ACM Computer Human Interaction}{Apr 30--May 6, 2022}{New Orleans, LA}
\acmBooktitle{CHI 2022: ACM Computer Human Interaction,
Apr 30--May 6, 2022, New Orleans, LA}
\acmPrice{15.00}
\acmISBN{978-1-4503-XXXX-X/18/06}

\usepackage[ruled, vlined]{algorithm2e}

\begin{document}

\title{Mouse Sensitivity Effects in First-Person Targeting Tasks}

\author{Ben Boudaoud}
\orcid{0000-0003-4195-0793}
\email{bboudaoud@nvidia.com}

\author{Josef Spjut}
\orcid{0000-0001-5483-7867}
\email{jspjut@nvidia.com}

\author{Joohwan Kim}
\email{sckim@nvidia.com}

\affiliation{%
  \institution{NVIDIA}
  \country{USA}
}
\renewcommand{\shortauthors}{Boudaoud et al.}

\renewcommand{\shorttitle}{Mouse Sensitivity in First-Person Targeting}

\begin{abstract}


Despite billions of hours of play and copious discussion online, mouse sensitivity recommendations for first-person targeting tasks vary by a factor of 10x or more and remain an active topic of debate in both competitive and recreational gaming communities.
Inspired by previous academic literature in pointer-based gain optimization, we conduct the first user study of mouse sensitivity in first person targeting tasks, reporting a statistically significant range of optimal values in both task completion time and throughput. 
Due to inherent incompatibility (i.e., lack of convert-ability) between sensitivity metrics adopted for prior pointer-based gain literature and those describing first-person targeting, we provide the first analytically demonstrated, statistically significant optimal sensitivity range useful for first-person camera controls. 
Furthermore, we demonstrate that this optimal sensitivity range arises (at least in part) from a speed-precision trade-off impacted by spatial task difficulty, similar to results reported in pointer-based sensitivity literature previously.
\end{abstract}




\maketitle

\section{Introduction}

\textit{Mouse sensitivity} is the relationship between physical motion of a mouse and the virtual motion this induces in a computer application.
This virtual motion could be translational (e.g., pointer-based tasks), rotational (e.g., first-person targeting), or even categorical (e.g., directional selection).
In modern computer systems mouse sensitivity is the product of several factors, including the mouse sensor resolution (up to the hardware limitations of the mouse) as well as the operating system's and application's sensitivity settings.

\begin{figure}
    \centering
    \includegraphics[width=\textwidth]{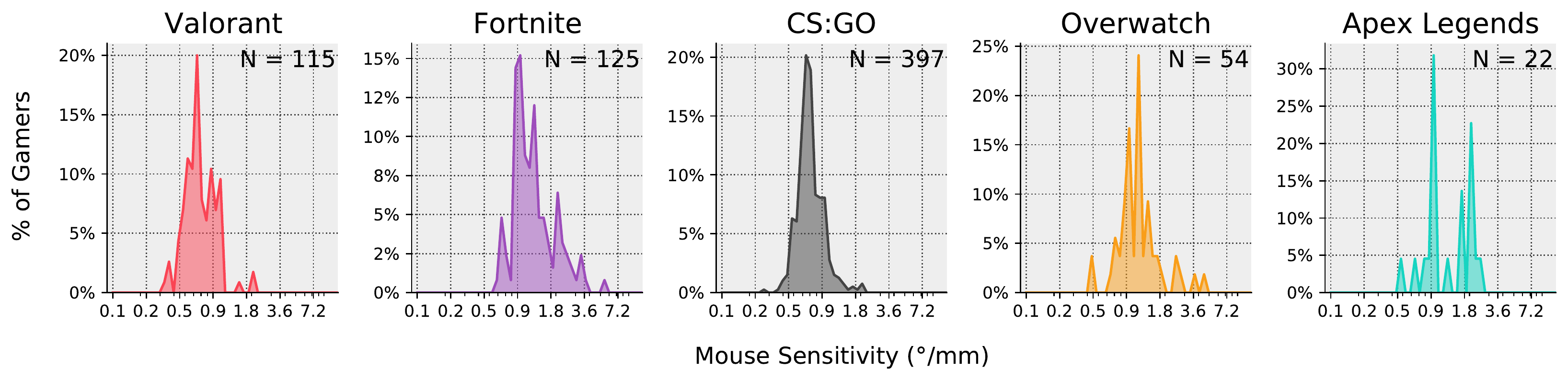}
    \vspace{-6mm}
    \caption{Professional player mouse sensitivities by game played (normalized per game). All games use the mouse for rotational camera motions. Note the logarithmic X-axis. The sensitivities are provided in cm/360$^\circ$, data sourced from \href{prosettings.com}{prosettings.com} in Sept. 2020.}
    \vspace{-2mm}
    \label{fig:ProSettingsSurvey}
\end{figure}

The competitive gaming community considers mouse sensitivity a significant factor of player performance in targeting tasks ~\cite{reddit2019mouse}.
Yet, there is wide variation in individual player settings. To demonstrate this, we collected and analyzed sensitivity settings data from \href{prosettings.com}{prosettings.com} for professional esports athletes across multiple First Person Shooter (FPS) gaming titles. 
The chart in Fig.~\ref{fig:ProSettingsSurvey} shows that the sensitivity range employed varies across professional gamers by up to 8x within a single title and by up to 12x across titles.
Furthermore, the default sensitivities for some of these games (scaled by common mouse DPI) lie beyond even the 12x range commonly used by professionals (see Table~ \ref{tab:default_cmp360}).


\begin{table}
    \centering
    \caption{Default $^\circ$/mm (cm/360$^\circ$) by game and DPI. 
    The game only sets the in-game value and the DPI depends on the mouse. Many gaming mice have 3200 DPI as a default while standard consumer mice tend to have 800 DPI by default.
    }
    \vspace{-2mm}
    \begin{tabular}{r|rrrrr}
        DPI & Valorant & Fortnite & CS:GO & Overwatch & Apex Legends \\
        \hline
        800 &  2.20 (16.34) & 4.90 (7.35) &  1.73 (20.78) & 3.12 (11.54) & 3.46 (10.39) \\
        3200 & 8.81 $\>\>$(4.08) & 19.60 (1.84) & 6.93 $\>\>$(5.20) & 12.47 $\>\>$(2.89) & 13.86 $\>\>$(2.60)\\
        
    
    \end{tabular}
    \label{tab:default_cmp360}
\end{table}

Academic research on mouse sensitivity has focused predominantly on productivity-oriented, pointer-based tasks, leaving a gap in research around first-person targeting. 
Many pointer-based studies recommend control-display (CD) gains of 2-8 ~\cite{pang2019effects, bohan2003gain, casiez2008impact, jellinek1990powermice, sandfeld2005effect, lin1992gain}, or over a factor of 4x in sensitivity range. 
This limited range makes sense, as pointer-based interaction is a reasonably uniform and repeatable task given similar user input/output devices and content mapping.
Unlike pointer-based user interfaces, designed for easing selection of static targets, FPS games present dynamic targets of various sizes at different virtual distances, deliberately challenging the player in order to reward  skill.
Thus, competitive FPS games eschew strict adherence to traditional user interface design patterns, as skilled gamers attempt to find the right mouse sensitivity to balance aiming speed and precision.

Though FPS targeting tasks have been demonstrated to conform to Fitts' law \cite{looser2005validity}, this does not imply that they share a common set of optimal sensitivities with pointer-based targeting tasks. 
Or, in fact, that the mouse sensitivities used to characterize pointer-based targeting sensitivity are convertible to first-person targeting sensitivities.
This paper presents the first study of mouse sensitivity's impacts on FPS targeting task performance, providing data useful for connection of pointer-based and first-person targeting mouse sensitivity in the future.



\section{Background}
Previous work has explored various aspects of spatiotemporal task difficulty, including the effects of mouse sensitivity on pointer-based targeting tasks. 
In this section we describe the related work foundational to our motivation and analysis.

\subsection{Sensitivity and Fitts' Law}
Fitts' law~\cite{fitts1954information} is a widely accepted model that explains spatial targeting and selection task difficulty as a function of target distance and size.
Nearly all formulations of Fitts' law and its variants \cite{fitts1954information,fitts1964information,mackenzie1992extending} normalize the distance to a target ($D$) by the target width $(W)$ in common units, producing a unitless difficulty ratio. The Index of Difficulty (ID), a measure of spatial task difficulty, is a base-2 logarithm of the D/W ratio (measured in bits), and is commonly linearly fit to task completion time. 
For ID computation in this work we use the Shannon formulation of Fitts' law as proposed by MacKenzie et al. \cite{mackenzie1992fitts} as provided in Eq.~\ref{eq:IDformulation}.

\begin{equation}
    ID = \log_2\left( \frac{D}{W}+1 \right) \mathrm{bits}
    \label{eq:IDformulation}
\end{equation}


\vspace{2mm}
Task \emph{throughput} \cite{mackenzie2008fitts}, also referred to as index of performance \cite{fitts1964information}, is the ratio of ID to task movement time ($T_{move}$), wherein changes in movement time resulting from human actuation are accounted for (see Eq.~\ref{eq:TPformulation}).

\begin{equation}
    Throughput = \frac{ID}{T_{move}}
    \label{eq:TPformulation}
\end{equation}
\vspace{0.5mm}

\begin{figure}
    \centering
    \includegraphics[width=0.7\textwidth]{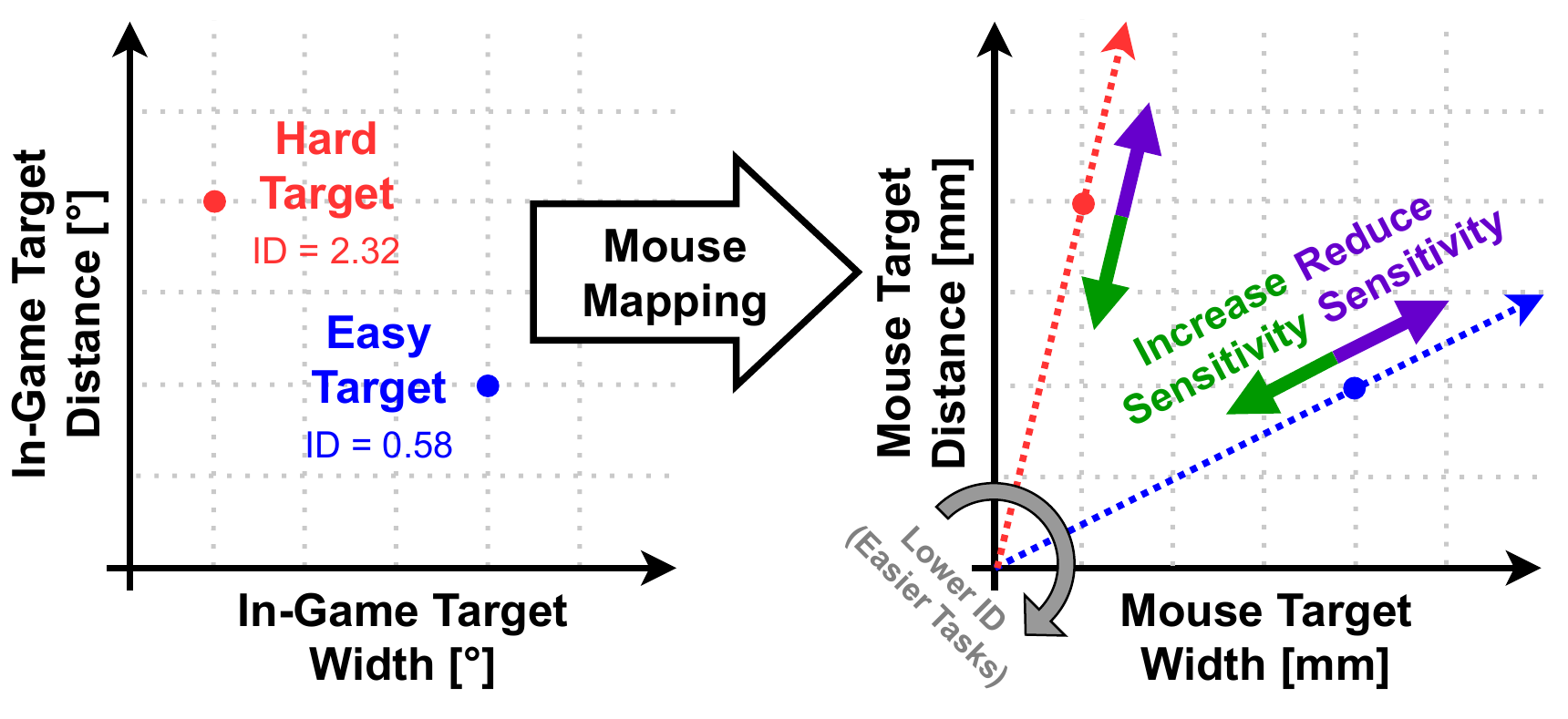}
    \caption{Plots demonstrating the mapping of an in-game, virtual rotational task to the (physical) mouse motion space under varying sensitivity, together with the orthogonality of sensitivity and task difficulty (ID).}
    \vspace{-4mm}
    \label{fig:MouseSensVsID}
\end{figure}

As Fig. \ref{fig:MouseSensVsID} demonstrates, changing mouse sensitivity either scales both $D$ and $W$ in physical mouse movement space or neither in in-game rotation space, and therefore does not modify the ID.
Instead, trends in optimal mouse sensitivity show up in task throughput.
Mouse sensitivity has been demonstrated to impact movement time, with previous work showing significant interaction with target size and task ID \cite{bohan2003gain,sandfeld2005effect}.
We explain the effect of mouse sensitivity on task performance using kinematic characteristics observed in our data in Section \ref{sec:SubmovementAnalysis}.

\subsection{Mouse Sensitivity in Pointer-based Targeting}
Most prior work in optimal mouse sensitivity has focused on pointer-based, Fitts' law targeting tasks. 
In these tasks users move the mouse,  translating an on-screen virtual proxy (cursor), towards a target, then click to indicate a selection. 
The control-display (CD) gain, or ratio between the travel distance of a controller (e.g., mouse) and the corresponding cursor movement on a display, has been the conventional choice of sensitivity metric in this work. Reported optimal CD gains vary from 2 to 16 (an 8x range) \cite{jellinek1990powermice,trankle1991factors,bohan2003gain,sandfeld2005effect,casiez2008impact} with some studies reporting higher optimal CD gain for tracking tasks \cite{senanayake2016pointing}. 

Older studies \cite{jellinek1990powermice,trankle1991factors} suggest lower CD gain, concluding mouse sensitivity does not significantly impact task completion time, with  quantization error limiting the range of sensitivity interesting for study. 
Later work \cite{lin1992gain,bohan2003gain,sandfeld2005effect,pang2019effects} implies that movement time does depend on mouse sensitivity, and highlights interactions with target size. Most recently, these results have been extended by monitoring limb usage range and rate of motion to show users do in fact scale their kinematic interaction with CD gain~\cite{casiez2008impact}. 
Additional studies indicate that tracking task completion time benefits from higher sensitivities or more direct methods of interaction like touch interfaces \cite{senanayake2016pointing}. 
Broadly speaking all surveys tend towards globally optimal CD gains in the range of 2-8, which aligns well with the default gains applied by operating systems and pointer-based user interfaces \cite{casiez2011no}.

\paragraph{CD Gain and Display Size/Distance}
Conventional CD gain, though a commonly reported sensitivity metric for pointer-based tasks, has a notable shortcoming in that the gain ratio takes into account the physical \emph{display}-space motion of a cursor. 
A large display viewed at a correspondingly large distance, such that it subtends the same visual angle as a common desktop monitor, will have a substantially higher CD gain than a perceptually equivalent desktop display.
That is, though the observed angular displacement of a cursor resulting from the same  user hand motion may be nearly identical between these cases, the CD gain is higher for larger displays as the cursor travels farther in display space \cite{casiez2008impact}. 
In spite of CD gain being incompatible with rotational FPS mouse sensitivity (see the following section), this display-contingent scaling of the metric means it is best applied to a relatively narrow range of display size, viewed at a consistent distance.

\subsection{Mouse Sensitivity in First Person Targeting}
\begin{figure}
    \centering
    \includegraphics[width=0.8\textwidth]{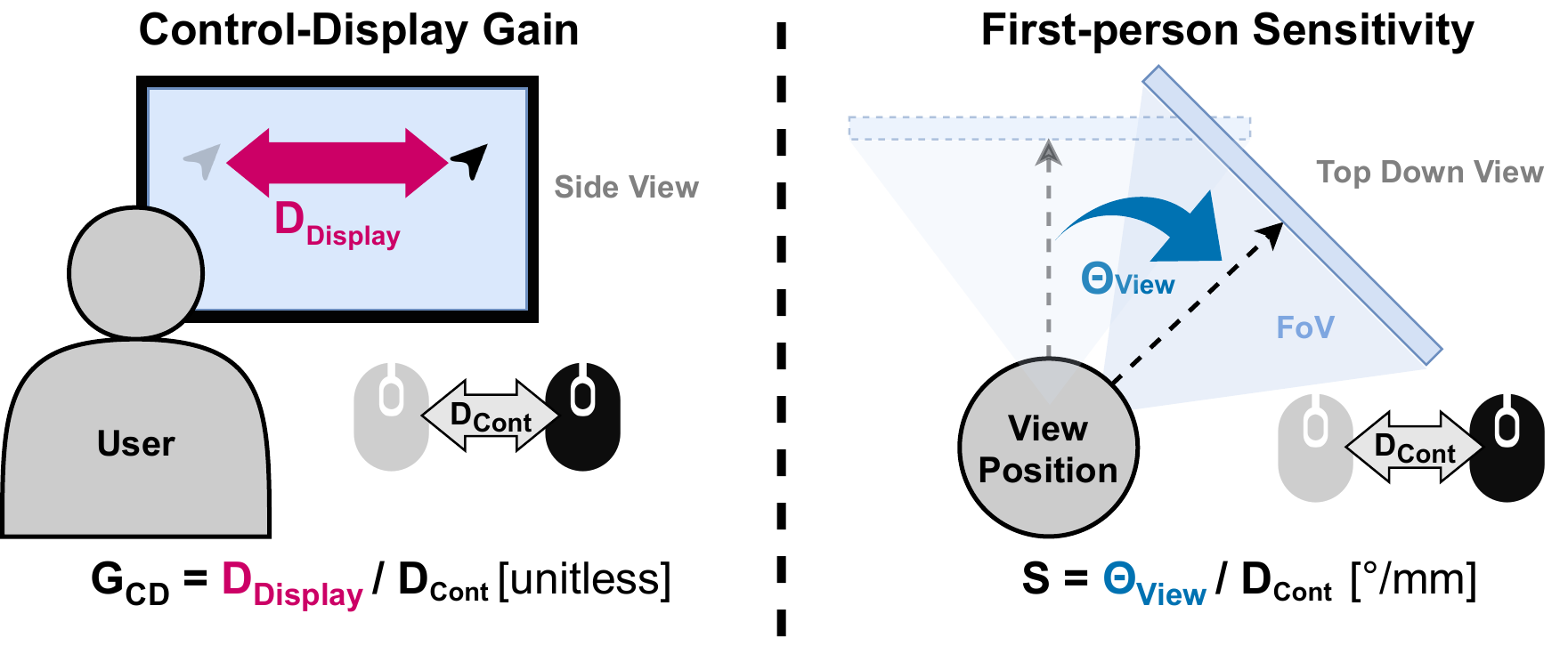}
    \caption{A comparison of CD gain and displacement-to-rotation sensitivity in their respective contexts (pointer-based and first-person targeting). CD gain uses common units of displacement (or velocity) to create a unit-less gain comparing mouse to pointer movement. Displacement-to-rotation sensitivity converts the same mouse movement into rotation of a view frustum about the current view position in the application's world space, resulting in a ratio that must have units.}
    \vspace{-6mm}
    \label{fig:CDGainvsRotation}
\end{figure}

First person targeting interfaces, such as those used in FPS games, provide an aim point indicator (often called the crosshair or \emph{reticle}) fixed at the center of the screen. 
A user rotates the view direction in azimuth and/or elevation by translating the mouse horizontally and/or vertically respectively, aligning a target with the reticle. 
Since the view direction rotates, the entire virtual world, including the target, moves in the direction opposite the mouse motion (e.g., mouse moves left, target moves right).
Similarly to pointer-based interfaces, the user clicks to perform selection.
Despite obvious differences from pointer-based targeting, both approaches share similar spatial task difficulty characteristics explained well by Fitts' law \cite{looser2005validity}.
The difference between pointer-based CD gain and first-person rotational mouse sensitivity is further demonstrated in Fig. \ref{fig:CDGainvsRotation}.

In the context of first-person aiming, many games provide settings sliders with the label "mouse sensitivity".
These sliders are typically unit-less, with higher values corresponding to faster view rotation per unit of physical displacement.
These slider's values (when reported) do not generalize across games and require a conversion factor to provide common units for gamers to communicate.
FPS players use a variety of metrics to talk about mouse sensitivity, the most universal of which is a displacement-to-rotation ratio, commonly measured in cm/360$^\circ$ (cm per full turn).
The primary motivation for this metric is ease of measurement in game by moving the mouse in one direction until a full rotation has been completed then measuring the distance the mouse traveled to complete that turn.
We use an inverse proportional metric, $^\circ/$mm (degrees per millimeter) to match the intuition that higher sensitivity means the aim moves more quickly per unit displacement.
For those who may be more familiar with the cm/360$^\circ$ metric, Table~\ref{tab:cmp360} maps the two for the settings used in this study.

\begin{table}[b]
    \centering
    \caption{Conversion between cm/360$^\circ$, cm/$^\circ$, and $^\circ/$mm mouse sensitivity settings selected for our study. The relationship between cm/360$^\circ$ and $^\circ/$mm is given by $y =36/x$ where $y$ is in $^\circ/$mm and $x$ is in cm/360$^\circ$.}
    \begin{tabular}{l|cccccc}
        \textbf{cm /} $\boldsymbol{360^\circ}$ & 160 & 80 & 40 & 20 & 10 & 5 \\
        \hline
        \textbf{cm /} $\boldsymbol{^\circ}$ & 0.444 & 0.222 & 0.111 & 0.056 & 0.028 & 0.014 \\
        \hline
         $\boldsymbol{^\circ}$ \textbf{/ mm} & 0.225 & 0.45 & 0.9 & 1.8 & 3.6 & 7.2 \\
    \end{tabular}
    \label{tab:cmp360}
\end{table}

\paragraph{Rotation To Displacement Sensitivity in Head-controlled Interfaces}
The sole prior result we are aware of in rotation-to-displacement sensitivity \cite{schaab1996comparison} focuses on head-controlled computer input devices, where a user's head rotation drives an on-screen pointer's translation, the inverse of translating a mouse to drive an virtual view rotation.
This work did find significant differences between different rotation-to-displacement ratios, but did not explore wide enough a range of sensitivities to identify a clear optimal region.
The range of sensitivity studied was likely relevant to the particular interface used in the demo.
The authors did observe that movement time decreased as they increased the rotation-to-displacement ratio over the range of 0.1-0.2$^\circ$/mm.
In addition, the authors note that while user's preference for a displacement-to-displacement gain changed with viewing distance, per the suggestion in the \emph{CD Gain and Display Size/Distance} paragraph above, subjects preference for rotation-to-displacement sensitivity did not significantly change with viewing distance.


\paragraph{CD Gain and First Person Targeting}
In addition to the problem of CD gain scaling with display size, there is no direct connection between a CD gain and the cm/360$^\circ$ or $^\circ/$mm sensitivity settings commonly used in FPS targeting applications. 
We could attempt to convert common historical CD gains into angular sensitivities by converting display space pointer motion (at an assumed distance) into angular motion instead.
However, if we were to convert a CD gain to an assumed $^\circ$/mm angular ratio, we would confound degrees of user focal point translation when tracking a pointer with degrees of camera rotation in mouse controlled cameras. 
This challenge is further complicated by common FPS practices, such as projecting a large rendered field of view (typically $\geq100^\circ$) onto a display that subtends a much smaller physical field of view (typically $<50^\circ$) and the user fixating at the center of the display instead of tracking a mouse cursor as it moves within the display area.

In an attempt to validate our results based on prior art measuring optimal CD gain in pointer-based tasks, we compare our experimental results to reported CD gains only using self-referential multiplicative ranges 
(i.e. the max to min ratio of 4x between a reported CD gain of 2 and 8 can be compared to that of a first-person sensitivity range of 0.5 and 2 $^\circ/$mm).
We avoid comparing reported CD gains between studies as different display size or viewing conditions may change the reported optimal values.

\paragraph{Non-linear Gain (Acceleration) in First Person Targeting}
Though relatively common in pointer-based interfaces \cite{casiez2008impact, casiez2011no, balakrishnan2004beating}, velocity-dependent sensitivity, often referred to as cursor/pointer acceleration or nonlinear CD gain, has been historically avoided in competitive FPS game play \cite{prosettings.acceleration, bestsettings2019accel, beaufort2020pointeraccel}. When applying nonlinear gain, mouse sensitivity is no longer a constant, but rather a (user-specified) function of the velocity at which the mouse is currently moving.

We suggest that competitive FPS gamers' aversion to nonlinear sensitivity is sourced largely from the added challenge it presents in actuation space, where skilled users often operate using an open loop targeting model to minimize task completion time. In this actuation model, players learn the spatial relationship between screen and cursor space, and move the mouse to align with a target and click as quickly as possible, without performing additional visual verification. This skill is often referred to as the \emph{flick shot} technique or simply \emph{flicking}.

When flicking, controlling the speed of view rotation change is more difficult as a movement takes place over a very short time period, and repeat-ability of the action becomes extremely important, as an additional perception-action cycle effectively negates the value of the technique over more traditional tracking approaches. For this reason, we suggest players prefer to simply learn to target their flick shots to the \emph{displacement} space of the mouse pad, explicitly avoiding the more precise speed control required to get benefit from nonlinear mouse sensitivity.

It is worth noting that some subsets of the gaming community have, and continue to, attempt to optimize various nonlinear sensitivity curves for various games of their choosing \cite{mamerow2020accelmyth, johnson2021accelCSGO}. 
However, the nonlinear sensitivity curves arrived at for these applications do not always conform to those suggested in prior academic literature or modern operating systems as they are deliberately intended to be easier to learn and adapt to quickly for open loop actions (i.e., 2-tier sensitivity structures).

\section{Experimental Design}
We designed an experiment to better understand the interaction of target width and distance with mouse sensitivity in a publicly available FPS experimentation platform~\cite{Spjut19FPSci}.
We provide our experiment configuration files along with this paper to enable others to repeat the study.

\subsection{Questions and Hypotheses}
The following questions motivate our experiment design.

\paragraph{Does a single optimal sensitivity (region) exist among competitive FPS gamers?}
We hypothesize that the optimal sensitivity regions observed in prior art result from humans' innate nonlinear kinematic characteristics, particularly those around excessively small targets and/or long travel distances.
If this is true, we should observe similar trends among competitive FPS gamers and more conventional (i.e. pointing) interface users despite years of aim training experience. 
To test this hypothesis, we included 6 fixed sensitivity conditions exponentially spaced over the set of sensitivities reported from our survey of esports athletes. 
We also widely varied target distance and size, into the ranges at which nonlinear motion characteristics have been observed in prior art.

\paragraph{Does an individual competitive gamers' preferred sensitivity setting approximate their generally optimal setting(s) for FPS targeting tasks?}
We hypothesize individual gamers' preferred sensitivity settings do tend to lie within regions of near optimal performance given sufficient time to adapt to in-game conditions.
To validate this hypothesis, we allocated a preferred sensitivity condition as the first session for all the participants. 
We re-used subjects' preferred sensitivity settings, selected through practice with a similar aiming task where the sensitivity was set by the subjects themselves.
Subjects were not made aware that their preferred sensitivities were among the tested conditions.
We compare subjects' performance at their preferred sensitivity to the surrounding sensitivities from our pre-selected set to evaluate this claim.

\paragraph{Does long(er) term muscle memory play a crucial role in aiming performance?}
Competitive gamers commonly assert that muscle memory can be "hard-wired" and their performance suffers when changing sensitivity too frequently.
We hypothesize that changing sensitivity (at least for the duration of  this experiment) does not have a significant impact on overall performance.
To test whether this is true, we allocate a preferred sensitivity condition as the final session of data collection, comparing these results to those from the first session.
By the time they start the last session, participants have experienced 6 widely varied sensitivity conditions (over a 32x range). 
This should considerably confuse muscle memory and, if it plays a crucial role, performance in the last session should be significantly worse than the first.


\subsection{Subjects}

15 experienced gamers (age 11-34, 1 female and 14 males) voluntarily participated  in our experiment. 
All subjects were regular FPS players with minimum play times of 10-20 hours per week. 
Subjects were not made aware of the experimental hypothesis. 
All subjects gave informed consent (for the two under-aged subjects, age 11 and 14 respectively, their parent gave informed consent). 
The experiment was conducted in accordance with the Declaration of Helsinki and each subject household was provided with their own dedicated system to minimize the need to clean/sanitize hardware resources between users.

One subject's test system experienced a degradation in performance (frame rate) during data collection, as a result we excluded this user from our data set. 
Two other subjects skipped a significant portion of (difficult) trials due to fatigue. 
Of these two subjects, one subject voluntarily repeated the experiment without skipping trials.
We chose to include this re-run subject data (shown in orange in all per-user plots) as much of our analysis was within subjects and, as the subject was already familiar with the experimentation platform and FPS gaming in general, we did not observe evidence of significant long-term training effects. 
We excluded the data from the other subject who skipped trials. 
In total, 13 out of 15 participating subjects' data are included in our analysis (age 11-33, 1 female and 12 males).

\subsection{Task and Stimulus}

\begin{figure}
    \centering
    \includegraphics[width=0.5\textwidth]{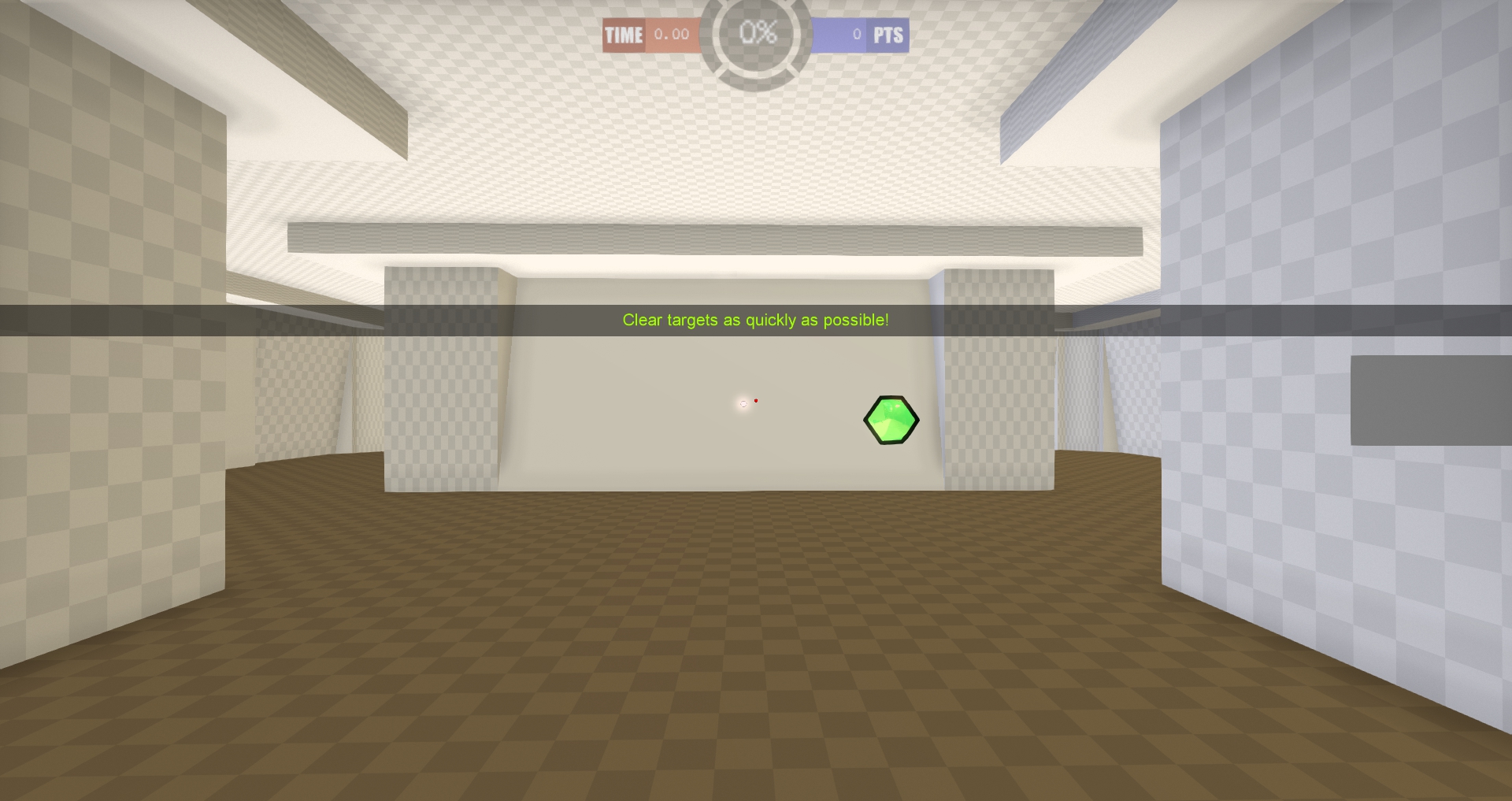}
    \hspace{1cm}
    \includegraphics[width=0.4\textwidth]{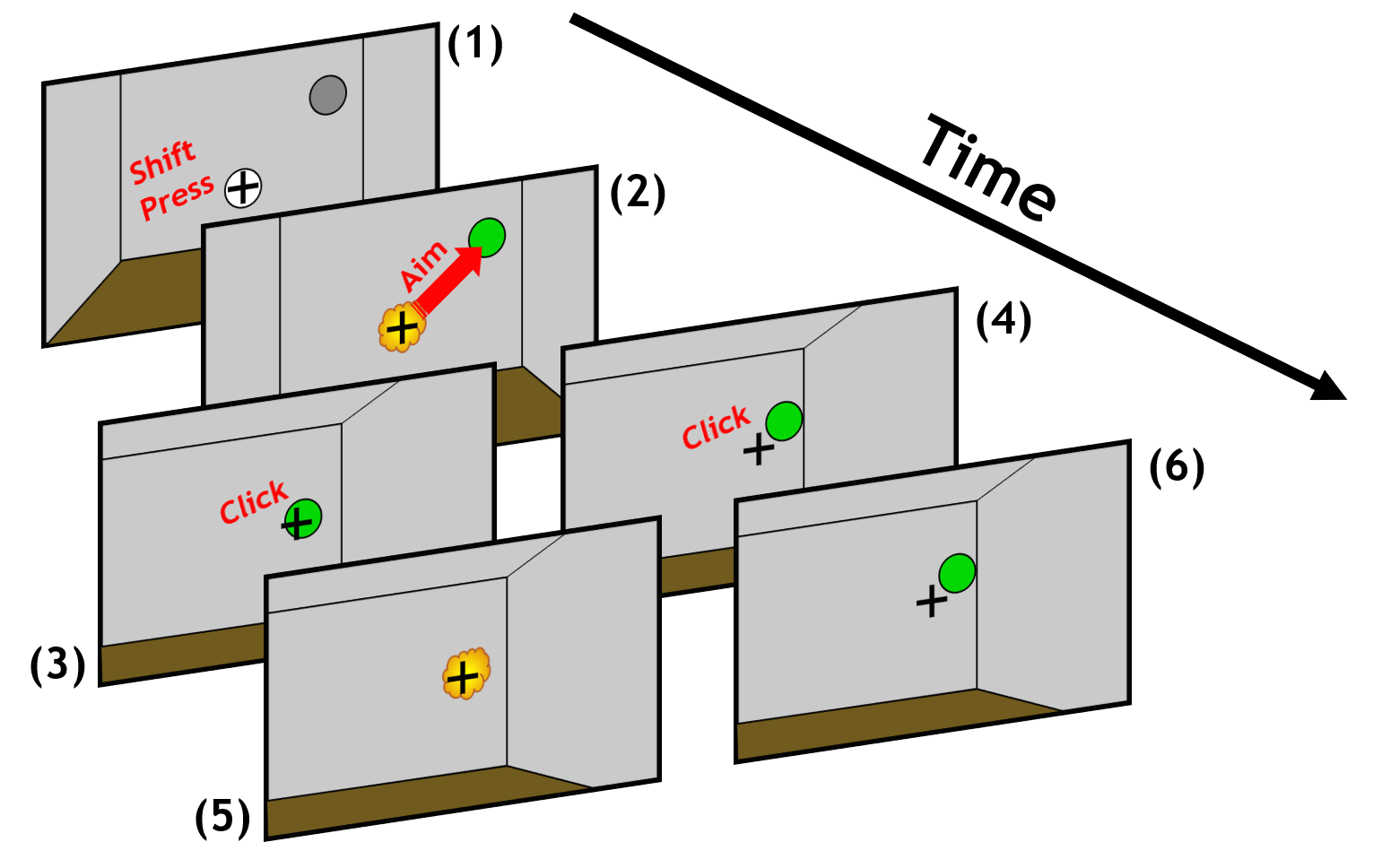}
    \caption{(Left) A screenshot from our application, the gray/white square at right was used to monitor system latency during the experiment. (Right) Flow of a single trial for our FPS task. (1) a reference target, shown in white, is used to recenter the user's view direction at the start of the trial, with the trial's test target displayed in gray to familiarize the subject with it's position. (2) after destroying the reference target the user tries to move their aim to the, now green, task target and click as quickly as possible. If the target is hit (3) the trial ends successfully (5), otherwise on a miss (4) the trial ends in failure (6).}
    \label{fig:TaskFlow}
    \vspace{-4mm}
\end{figure}

Subjects performed a first-person perspective aiming task with a stationary target (as shown in Fig. \ref{fig:TaskFlow}). 
We avoid rendering a weapon or complex background scene to minimize target view occlusion or possible distractions while subjects performed the task.

A trial began with presenting a stationary, white reference target 0.57$^\circ$ in diameter (8 pixels on the screen) and a stationary, but inactive, grey test target preview. 
Subjects aimed by translating the mouse to align their reticle with the reference target. 
The task timer started as the subject cleared the reference target by pressing the left shift key on the keyboard, activating the test target (indicated visually by changing its color to green). 
We chose to use left shift to begin the trial as this avoided effects from multi-purposing the left mouse button for both starting and ending the trial, including accidentally skipping trials by clicking twice.
The subject then quickly changed their aim by moving the mouse toward the test target. 
The task timer stopped when the subject clicked the left mouse button, with only a single click allowed per trial. 
If the reticle was aligned with the target when this click occurred the  task is completed successfully, otherwise the task ends in failure. 
Because the test target was visible before the timer started, the measured time consisted only of motor response time (visual search time and initial reaction time were excluded).

Subjects were instructed to clear the test targets as quickly and accurately as possible, and provided with a score indicative of this metric that was updated per trial. 
If the reticle was within the target bounds at the time of click, then subjects were scored by the amount of remaining task completion time in seconds (faster completion implies higher score). 
In the case of a failure, or the reticle being outside of the target at the time of click, the maximum completion time (1.5 s) was assumed for scoring purposes, resulting in a score of 0. 
The score penalty of the maximum completion time was substantially larger than that of the average task completion time (typically 0.4 - 0.8 s).
This scoring system is intended to motivate the subjects to maintain relatively high accuracy levels throughout the experiment (average hit rate of >85\%). 

\subsection{Conditions}

Each subject completed 8 sessions of 500 trials, with one mouse sensitivity level assigned to each session. The first and last sessions used the subject's preferred sensitivity setting. We varied target distance and size within each session to provide targets with a wide range of index of difficulties (see Table \ref{tab:IDRange}).

\subsubsection{Sensitivities}

Sensitivity was the main variable of interest in our study. We selected 7 sensitivity levels for exploration: 0.225, 0.45, 0.9, 1.8, 3.6, 7.2 $^\circ/$mm selected for their round number in the cm/360$^\circ$ scale (Table~\ref{tab:cmp360}), 
as well as the preferred sensitivity of each subject. We controlled sensitivity by changing the scaling multiplier between the movement of the mouse and the rotation of the rendering camera in the experimentation platform. 

\subsubsection{Target Width and Eccentricity}
In each condition we provided targets that varied both in width and eccentricity (i.e., distance from the central view direction, where the reference target was located).
The targets were all icosahedra with diameters of: 0.57$^\circ$, 1.15$^\circ$, 2.29$^\circ$, 4.59$^\circ$, and 9.15$^\circ$, expressed in perspective angle from the player view camera, which used a 103$^\circ$ rendered field of view. 
When the targets are projected near the center of the screen (25 inch or 635 mm, 1920x1080 monitor), these targets were 8, 15, 31, 61, and 122 pixels wide. 

\begin{table}
    \centering
    \caption{Index of difficulty range (in bits) as computed by Equation \ref{eq:IDformulation}) for our selected target width and distance pairings. All targets were within the range of 0.83-5.50 bits of difficulty with an equal number of targets drawn from each range of conditions described below.}
    \begin{tabular}{c|c|c|c|c|c|}
        \textbf{Target Distance} & \multicolumn{5}{c}{\textbf{Target Width}} \\
        \textbf{(Azimuth x Elevation)}&  9.15$^\circ$ & 4.59$^\circ$ & 2.29$^\circ$ & 1.15$^\circ$ & 0.57$^\circ$ \\
         \hline
          7-9$^\circ$ x 0-1$^\circ$ & 0.83-0.99 & 1.35-1.57 & 2.03-2.31 & 2.84-3.15 & 3.74-4.08 \\
          \hline
         9-11.7$^\circ$ x 0-1.3$^\circ$ & 0.99-1.19 & 1.57-1.84 & 2.30-2.62 & 3.14-3.49 & 4.07-4.44 \\
         \hline
         11.7-15$^\circ$ x 0-1.7$^\circ$ & 1.19-1.41 & 1.83-2.10 & 2.61-2.92 & 3.48-3.82 & 4.43-4.78 \\
         \hline
         15-19.4$^\circ$ x 0-2.2$^\circ$ & 1.40-1.65 & 2.10-2.40 & 2.92-3.25 & 3.81-4.17 & 4.77-5.14 \\
         \hline
         19.4-25$^\circ$ x 0-2.8$^\circ$ & 1.64-1.91 & 2.39-2.70 & 3.24-3.58 & 4.16-4.52 & 5.13-5.50 \\
         \hline
    \end{tabular}
    \label{tab:IDRange}
\end{table}

The target distance from the central view direction randomly varied between 7-25$^\circ$ horizontally and 0-2.8$^\circ$ vertically, with this distance range logarithmically divided into five sub-regions for a more uniform distribution of index of difficulty (Table \ref{tab:IDRange}). Each trial's target distance was drawn uniformly from the sub-region designated for that trial. The total distance variation corresponded to 93-339 pixels from the central view direction. The combination of selected target distance and width resulted in a range of ID of 0.83-5.50.


\subsubsection{Equipment}
All subjects used their own copy of an identical PC setup (Intel Core i7-9700k @ 3.6 GHz, 32 GB of RAM, RTX 2080 Ti). Subjects were seated about 24 inches (610 mm) from a 240 Hz, 1920 x 1080 25 inch (635 mm), LCD monitor. All subjects completed the experiment with the same Logitech G203 mouse at a 3,200 DPI sensor resolution, which was sufficiently high to provide pixel-level rotations in all tested sensitivity conditions, preventing any impact of quantization error.

\subsection{Procedure}

All subjects had experience with the experimentation platform prior to this study, acquainting them to the visual environment, controls, and high-level targeting task.
No further orientation was needed beyond clarification on the scoring system used in this experiment. 
To encourage all subjects to put in their best effort, we issued a \$25 gift card to the subject with the highest score after completing all 8 sessions.

Each session consisted of 500 trials (4000 trials/subject total), which were further divided into 10 blocks of 50 trials. 
Subjects were encouraged to take breaks between blocks to refresh themselves as needed. 
Subjects could complete as many sessions as they liked in a single sitting, but were strongly encouraged to stop if they felt any mental or physical fatigue. 
Each session took about 10 minutes to complete, with the entire experiment lasting about 80 minutes per subject. 
All participants completed the experiment within a single week.

\section{Results and Analysis}
We analyze our results from a mean and variation-based perspective, as all of these metrics impact player performance in the competitive gaming context. We consider hit rate, task completion time, and task throughput (using a Fitts' based ID formulation). 

We discard the first 50\% of trials from each session as an adaptation period for the new sensitivity condition, leaving 250 trials per session (see Section \ref{sec:Discussion} and Fig. \ref{fig:TrainingEffects} for more information on average adaptation rate), or 2,000 trials per user, for analysis.
Not all trials were successfully completed, in some cases the user clicked with the reticle outside of the target bounds.
We choose to analyze unsuccessful trials along with successful ones based on the assumption these trials were still valid aiming attempts.
Accordingly, we discard trials in which the aim error at the time of click was greater than 50\% of the initial distance to the target. 
These are trials where the user translated the view less than halfway to the target before clicking. 
45 total trials were removed based on this criteria, accounting for <0.2\% of all data and <0.5\% (10/2000) of any given user's trials.


\subsection{Hit Rate}
\begin{figure}
    \centering
    \includegraphics[width=0.49\textwidth]{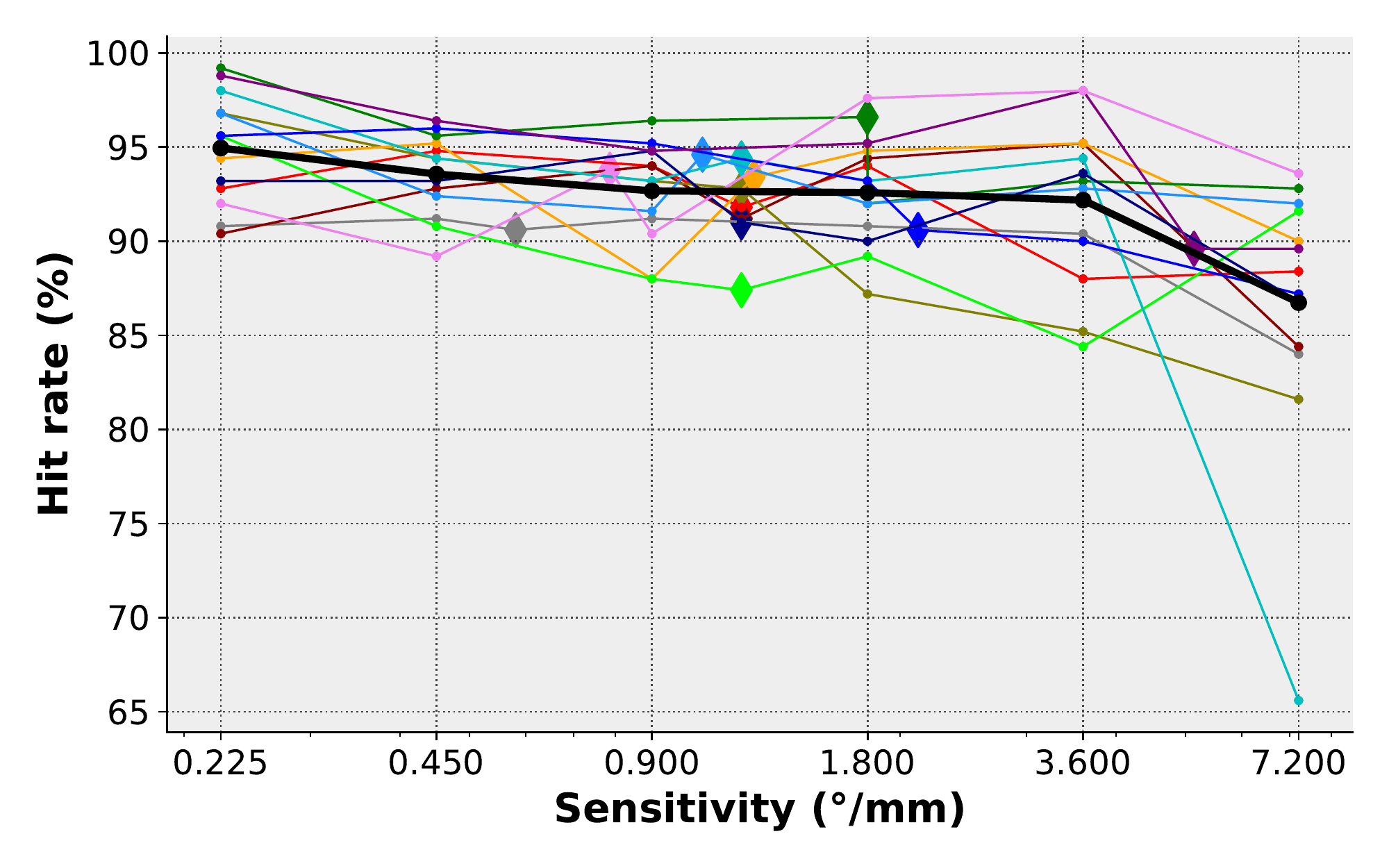}
    \vspace{-4mm}
    \caption{Subject trial hit (success) rate vs sensitivity per user (colors) and overall (thicker black line). Sensitivity preferences are displayed as diamonds. Note the X-axis is logarithmic.}
    \label{fig:HitRate}
    \vspace{-4mm}
\end{figure}

We analyze our data for hits as a percentage of total trials, where each trial is either a hit or miss. Only one shot was allowed per trial. 
As expected, most of our subjects maintain a hit rate above 85\% for nearly all sensitivity conditions.
At the highest sensitivity setting nearly all users' hit rate decreased.
Unlike other analyzed metrics, hit rate does not show a strong optimal point overall or per user in the range tested.
Generally speaking hit rate was highest for the lowest sensitivities, with minimal change across all but the highest sensitivity tested.
Though ANOVA reports a significant main effect ($F(5,72) = 6.12$, $p < 0.0001$) of sensitivity on hit rate for the data in Fig. \ref{fig:HitRate}, pairwise t-testing reveals differences of the means to be significant only when comparing the highest sensitivity setting (7.2$^\circ/$mm or 5 cm/360$^\circ$)  to the remainder of sensitivities ($p = 0.0146$). 

\begin{figure}[b]
    \vspace{-4mm}
    \centering
    \includegraphics[width=0.48\textwidth]{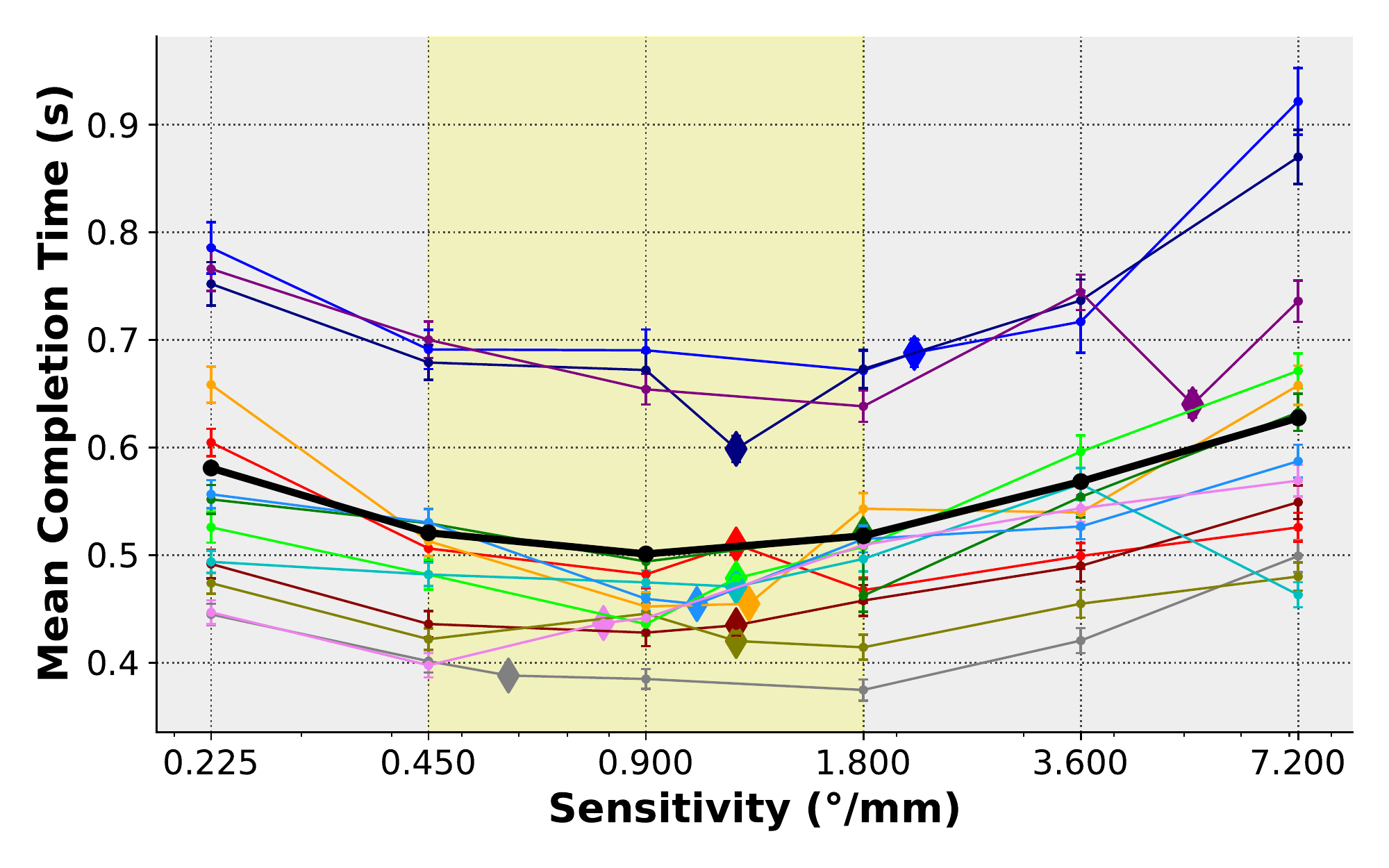}\hspace{3mm}
    \includegraphics[width=0.48\textwidth]{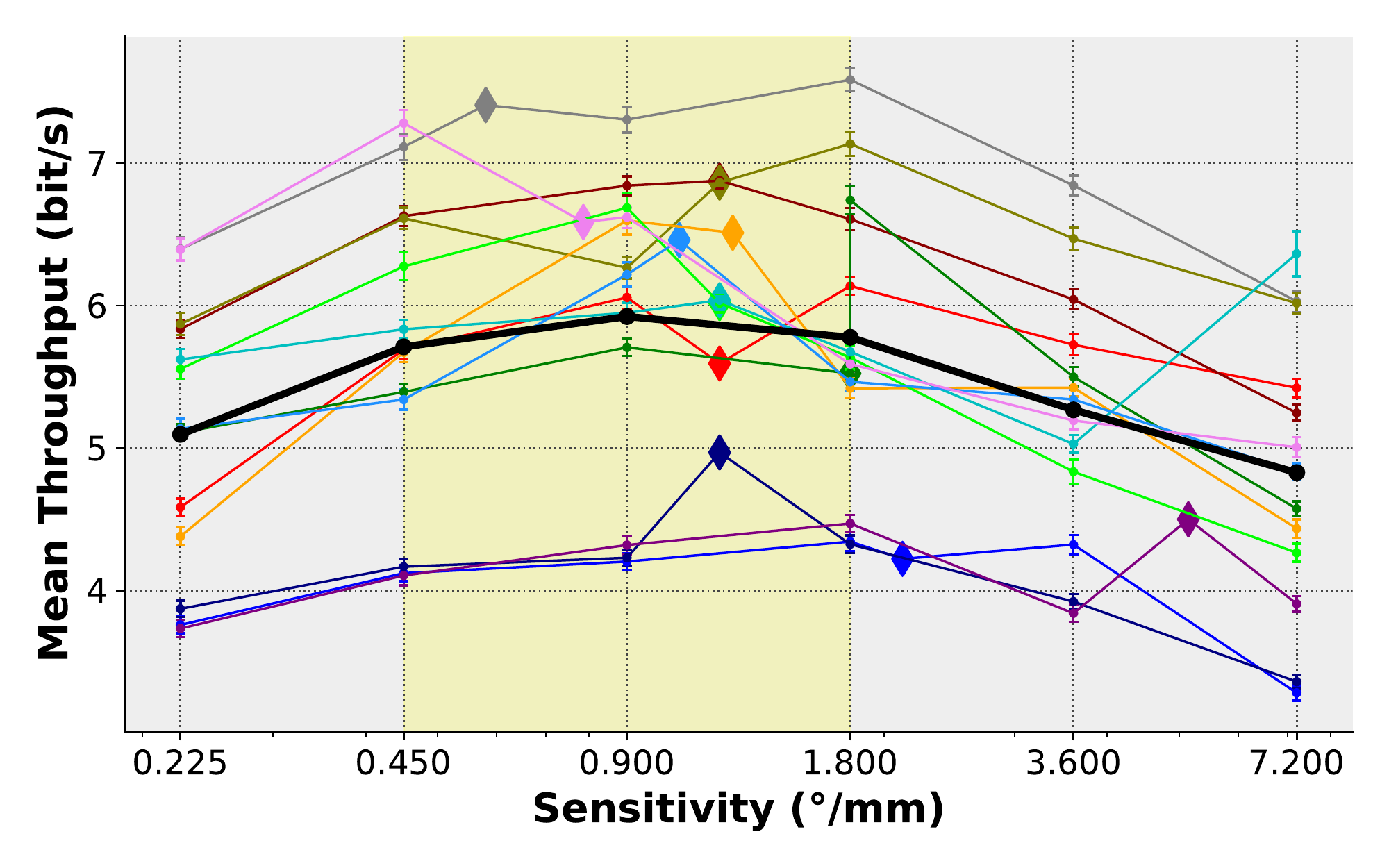}
    \vspace{-4mm}
    \caption{Mean task completion time (left) and throughput (right)  per user (colors) and overall (thicker black line) versus sensitivity. Per-user preferred sensitivity conditions are indicated by the diamond markers. Error bars represent the standard error metric. The X-axis is logarithmic and the yellow shaded region represents the zone of optimal performance.}
    \label{fig:MeanTaskTimeThroughput}
\end{figure}

\subsection{Task Completion Time and Throughput}
Within and between subjects we observed strong evidence of effects of mouse sensitivity on task completion time and throughput. Mean task completion time and throughput versus sensitivity are plotted per user (one per color) and overall (thicker black line) in Fig. \ref{fig:MeanTaskTimeThroughput}. Each user's preferred sensitivity condition is indicated by a diamond marker, with many of the users selecting an overlapping setting of 1.2$^\circ/$mm (30 cm/360$^\circ$).

ANOVA indicates a significant main effect of sensitivity on task completion time ($F(5,19955) = 111.72$, $p < 0.0001$), with pairwise t-testing validating these differences ($p < 0.001$) for all adjacent sensitivity pairings except those in the 0.45-1.8 $^\circ/$mm (20-80 cm/360$^\circ$) range. The range of this optimal zone mirrors the factor of 4 in optimal mouse sensitivities reported for pointer-based tasks in prior art.

In addition to completion time, we analyze mean task throughput (see Eq. \ref{eq:TPformulation}) per individual and over all subjects based on our experiment data. 
This metric decorrelates the spatial difficulty of a particular task from the time measured for completion of that task by taking the ratio of the index of difficulty (in bits) to the completion time.
ANOVA confirms a significant main effect of mouse sensitivity on throughput ($F(5,19955) = 275.59$, $p < 0.0001$). Additional pair-wise t-testing  between adjacent sensitivities confirms significant differences of mean ($p < 0.001$) for all tested sensitivities.

Interestingly, many subjects performed better at a pre-selected sensitivity than at their preferred sensitivity setting, both in completion time and task throughput, though generally speaking this preferred sensitivity was within their range of optimal performance. 
All users performed optimally in mean performance in the range of 0.45-1.8$^\circ$/mm (20-80 cm/360$^\circ$).
One user (shown in dark green) selected a preferred sensitivity setting that overlapped with a selected tested sensitivity (1.8$^\circ/$mm or 20 cm/360$^\circ$). 
This accounts for the jump in the mean task completion time and throughput for this user at the setting.
This discontinuity likely occurred due to fatigue impacting the session that overlapped with their preferred sensitivity conditions (the final session of that day for the user), but not the preferred sensitivity sessions themselves, as these were completed on other days.

\subsection{Variance of Task Completion Time}
In addition to mean task completion time and throughput, esports athletes seek to minimize variation of task completion time. In many cases increasing repeatability (reducing the spread) of a performance metric may in fact be more vital to overall outcomes than producing a small change in the mean of the metric, as it makes individual performance more predictable and team strategy more robust. For this reason we choose to analyze the standard deviation of task completion time per subject as an additional measure of optimality of sensitivity settings.

\begin{figure}
    \centering
    \includegraphics[width=0.47\textwidth]{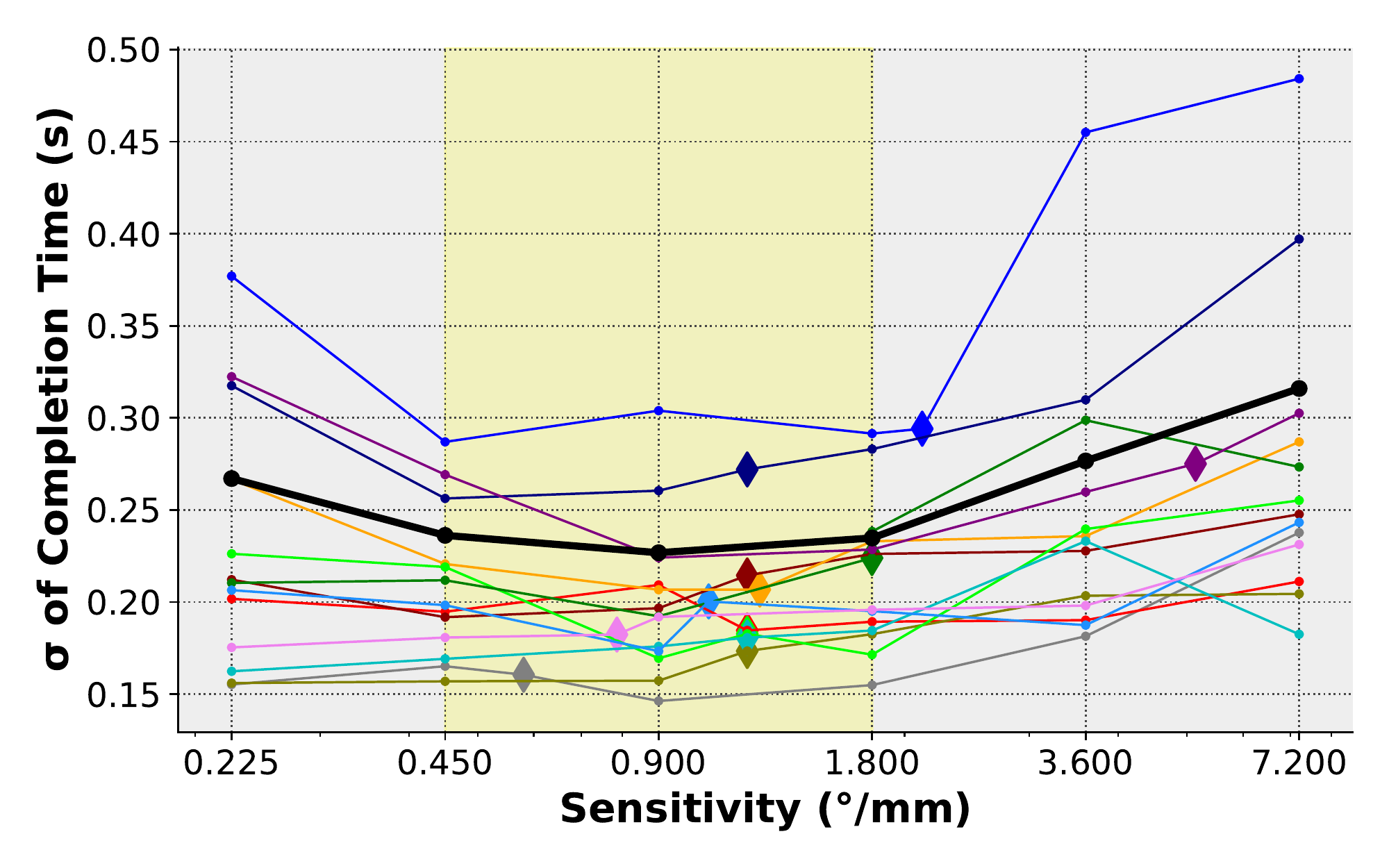}\hspace{7mm}
    \includegraphics[width=0.47\textwidth]{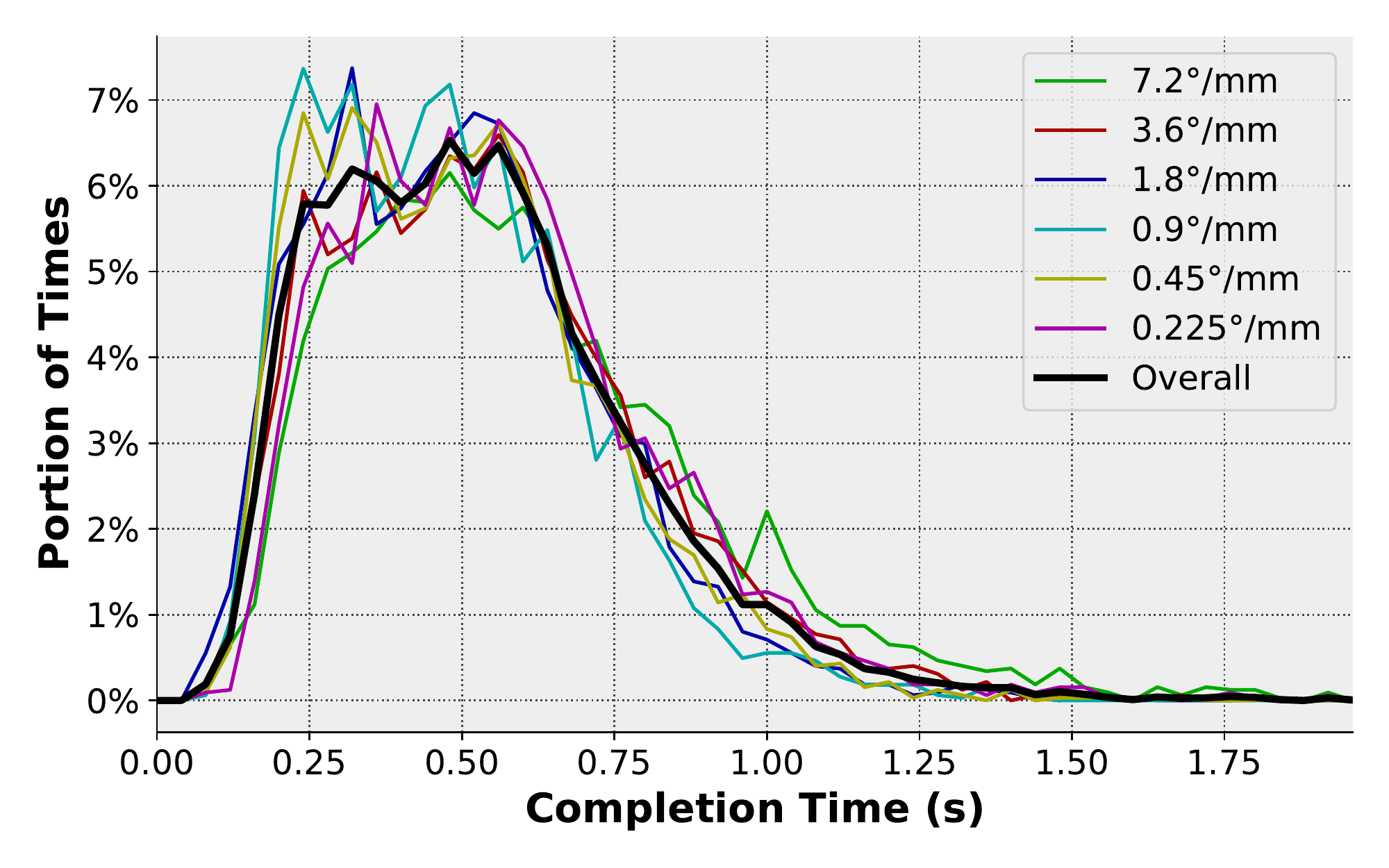}
    \vspace{-3mm}
    \caption{(Left) standard deviation of task completion time by user (one per color) and overall (thicker black line) for completed trials. Diamond markers indicate preferred sensitivity conditions. The X-axis is logarithmic and the yellow shaded region represents the zone of optimal performance. (Right) per sensitivity and overall distributional shapes for task completion time.}
    \label{fig:StdDevTaskTime}
    \vspace{-4mm}
\end{figure}

Figure~\ref{fig:StdDevTaskTime} shows the standard deviation of task completion time by subject versus mouse sensitivity, color coded the same as the mean time and throughput in Fig. \ref{fig:MeanTaskTimeThroughput}. 
We also present the distributional shape of completion time across subjects for each sensitivity condition.
Again, we see optimal (minimum) variance for mouse sensitivities in the range of 0.45-1.8$^\circ$/mm (20-80 cm/360$^\circ$), agreeing with the optimal range from our mean completion time and throughput results. 
Note that though we use standard deviation as a metric to analyze the spread of this data it is not distributed according to a symmetric Gaussian distribution (see Fig. \ref{fig:StdDevTaskTime}).

To validate whether the differences in standard deviation/variance reported above were significantly different between sensitivities we perform a Levene test for equality of variances. Overall results indicate a significant difference in task completion time variance between sensitivities ($W = 44.12$, $p < 0.0001$). Additional pairwise Levene testing between adjacent sensitivities yields significant differences ($p < 0.005$) for all sensitivities outside of the 0.45-1.8$^\circ$/mm (20-80 cm/360$^\circ$) range.




\subsection{Submovement Analysis}
\label{sec:SubmovementAnalysis}
The previous analysis in this section, specifically that of task completion time and throughput, begs an interesting kinematic question.
What was the effect of sensitivity on physical user motion?
Did users exploit higher sensitivities to turn faster in-game at the same real-world motion velocity, and if so, why does performance often degrade more at the highest sensitivities than the lowest ones?
To answer these questions we turn to submovement analysis.

Submovements are short, (pseudo)ballistic motion trajectories that arise from the human motor response planning process \cite{meyer1988optimality, chen2015structure, chua1993visual,hsieh2017submovement}. 
We employ a submovement segmentation technique inspired by that of Meyer \cite{meyer1988optimality}. 
We extend this work by considering periods of initialization, pauses, and verification \cite{chen2015structure} and online regulation of movements \cite{chua1993visual,hsieh2017submovement}.

\begin{algorithm}[b]
    \SetAlgoLined
    \caption{Submovement Parsing Algorithm}
    \KwData{$time$, azimuth ($az$), elevation ($el$) series}
    \KwResult{list of submovement start/end times ($move_{start}, move_{end}$)}
    
    $az_{filt},el_{filt} = filter_{LP}(time,az,el)$\;
    $v = |(\delta(az_{filt}), \delta(el_{filt}))|/\delta(time)$\;
    \For{t in time}{
        \uIf{previously in a submovement}{
            \lIf{$v[t] < V_{end}$ \textbf{and} $t-t_{start} > T_{min}$}{
                $move_{end}.add(t)$
            }
        }
        \uElseIf{$v[t] > V_{start}$}{
                $move_{start}.add(t)$; $t_{start} = t$\;
        }
    }
\label{alg:submovement}
\end{algorithm}

\begin{figure}
    \centering
    \includegraphics[width=0.7\textwidth]{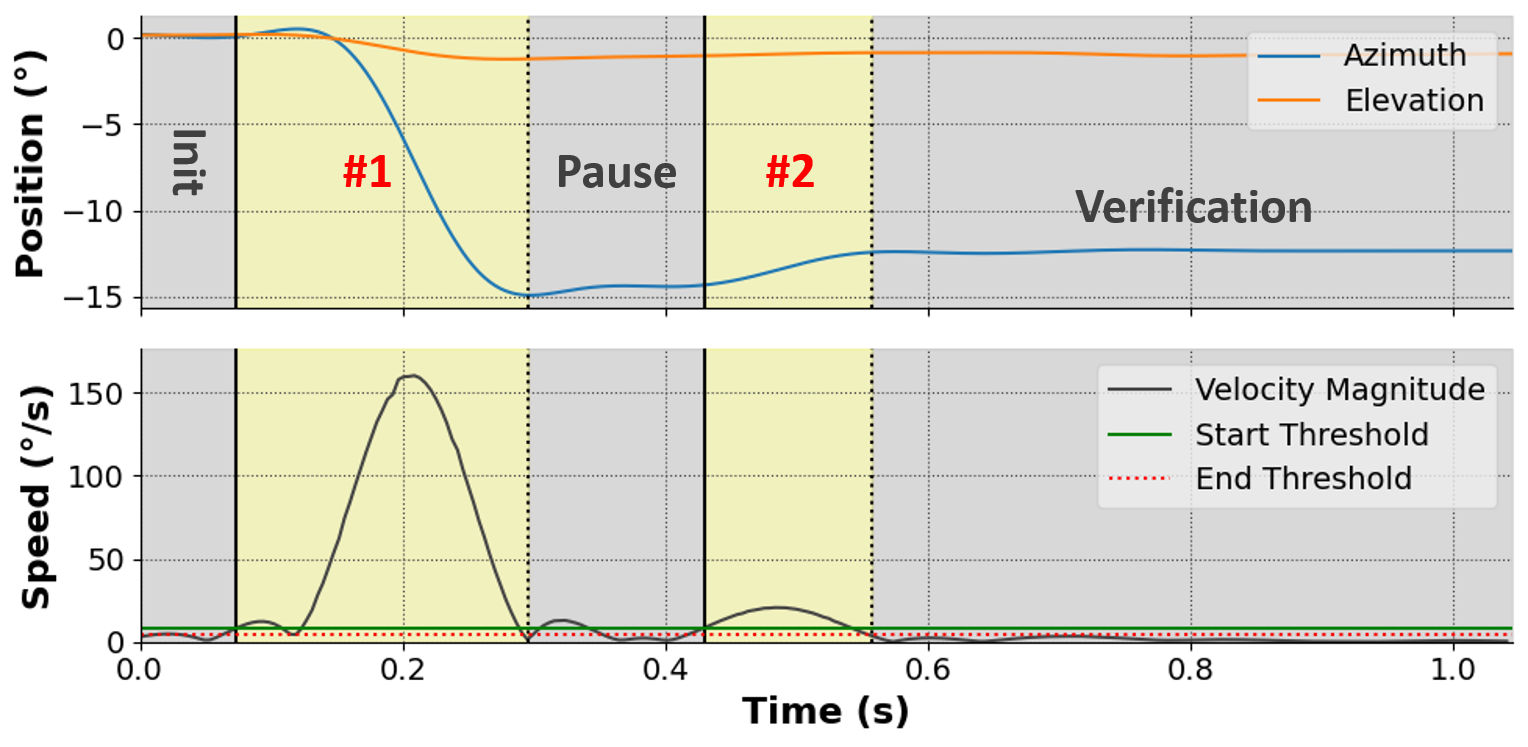}
    \vspace{-4mm}
    \caption{An example of a partitioned set of submovements with initialization, pause, and verification periods (shown in gray), together with 2 distinct submovements (shown in yellow). The bottom plot shows the velocity vector magnitude of the azimuth/elevation time series reported in the top plot along with thresholds for detection.}
    \label{fig:ExampleSubmovement}
    \vspace{-2mm}
\end{figure}

Algorithm \ref{alg:submovement} outlines the approach used for parsing submovements for this work. 
A 5th order, 7 Hz low pass Butterworth filter is used to smooth the azimuth/elevation data prior to submovement processing. 
Initialization, pauses, and verification periods are distinguished from periods of motion using the start velocity threshold ($V_{start}$). 
Multiple submovements detected within a single psychological refractory period ($T_{min}$) are considered one submovement. 
A submovement ends whenever the minimum duration of a submovement ($T_{min}$) has elapsed and the combined angular velocity falls below the end threshold ($V_{end}$).
For our analysis we select $V_{start} = 8^\circ/$s, $V_{end} = 4^\circ/$s, and $T_{min}=80$ms.
These values are selected empirically based on manual validation of the segmentation they produced, with $T_{min}$ approximating lower bound estimates of the psychological refractory period.
An example of submovements segmented from a single trial's azimuth/elevation aim time series using these values is provided in Fig.~\ref{fig:ExampleSubmovement}, demonstrating an initialization, pause, and verification window together with 2 segmented submovements.

\begin{figure}
    \centering
    \includegraphics[width=0.32\textwidth]{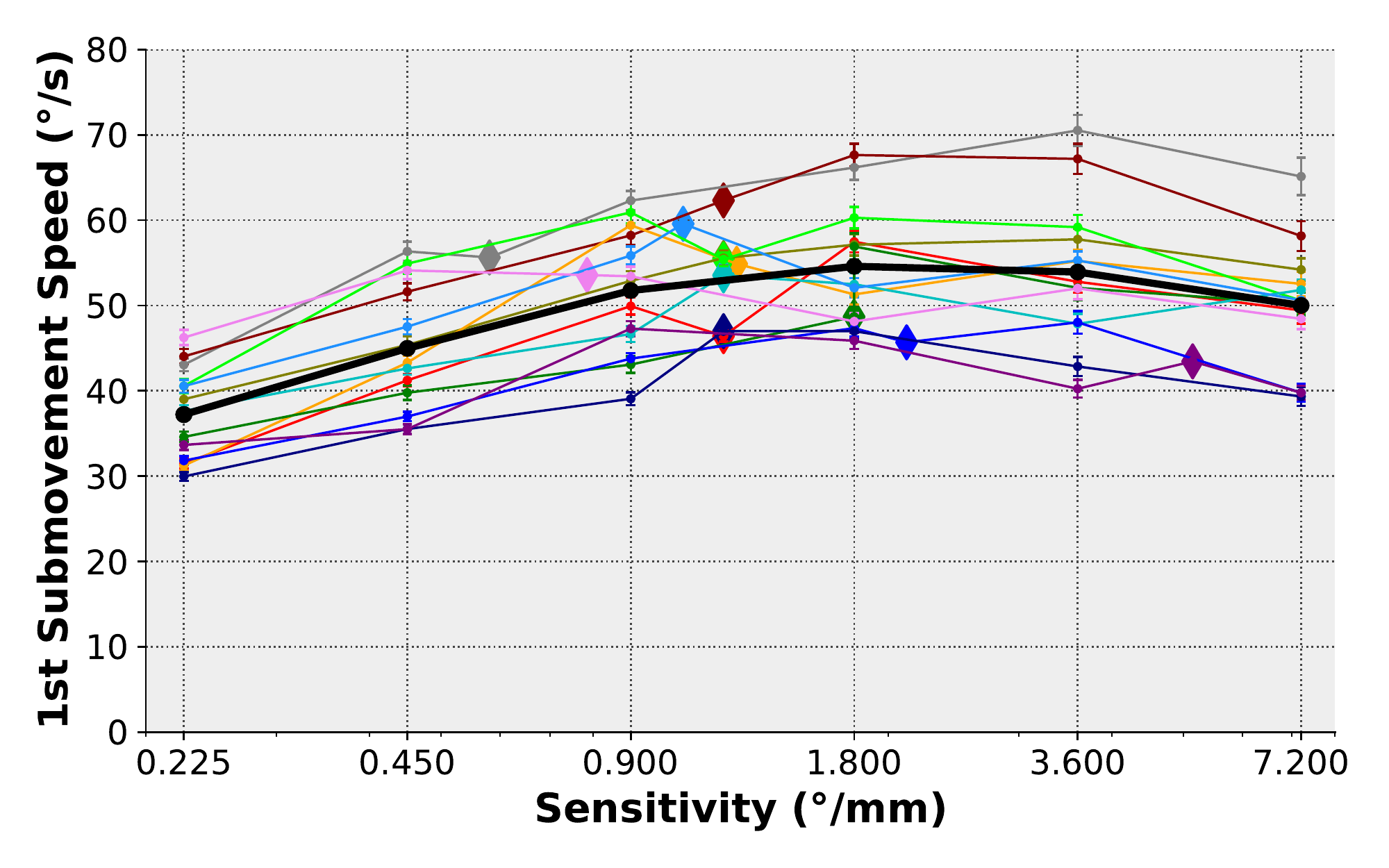}
    \includegraphics[width=0.32\textwidth]{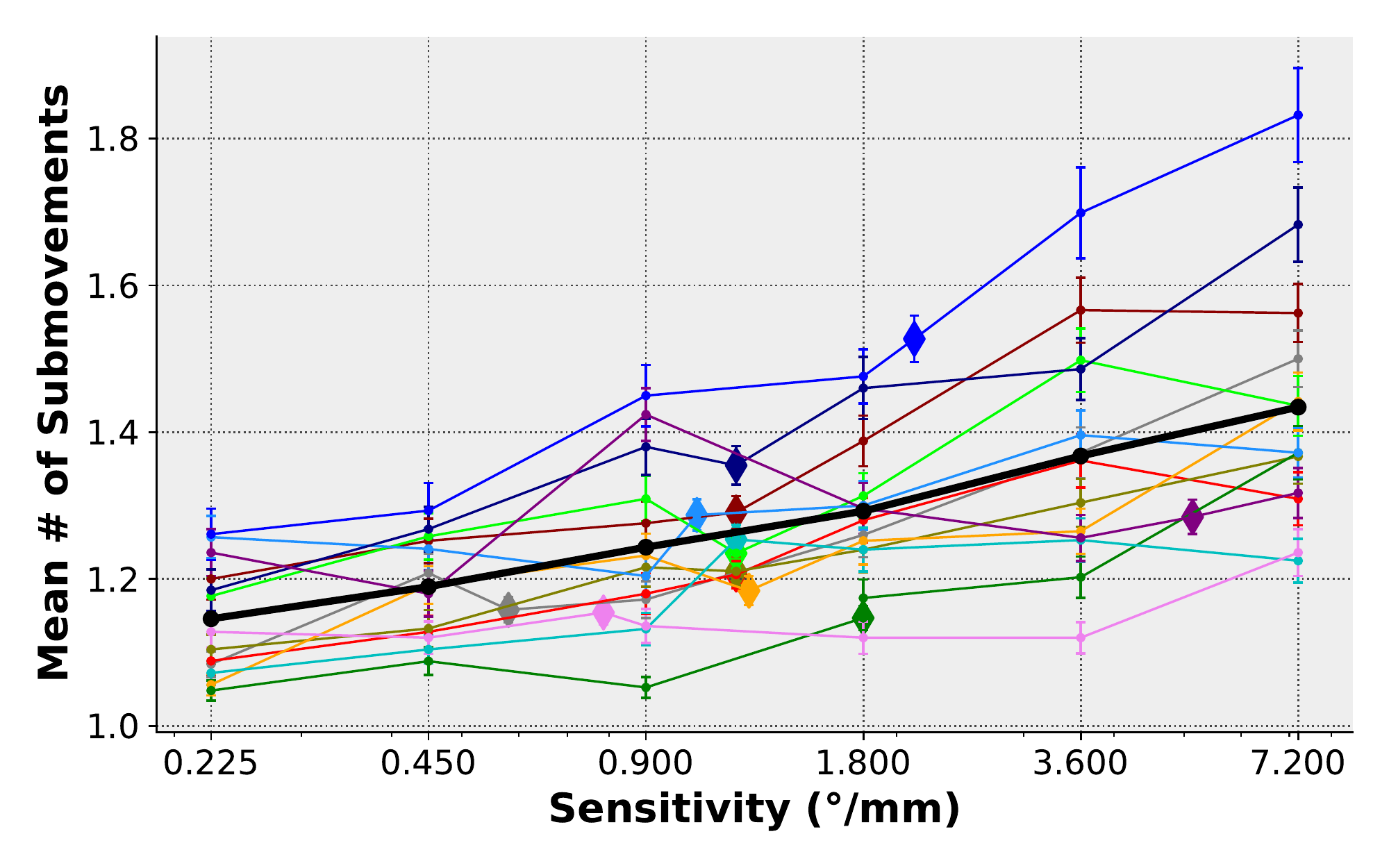}
    \includegraphics[width=0.32\textwidth]{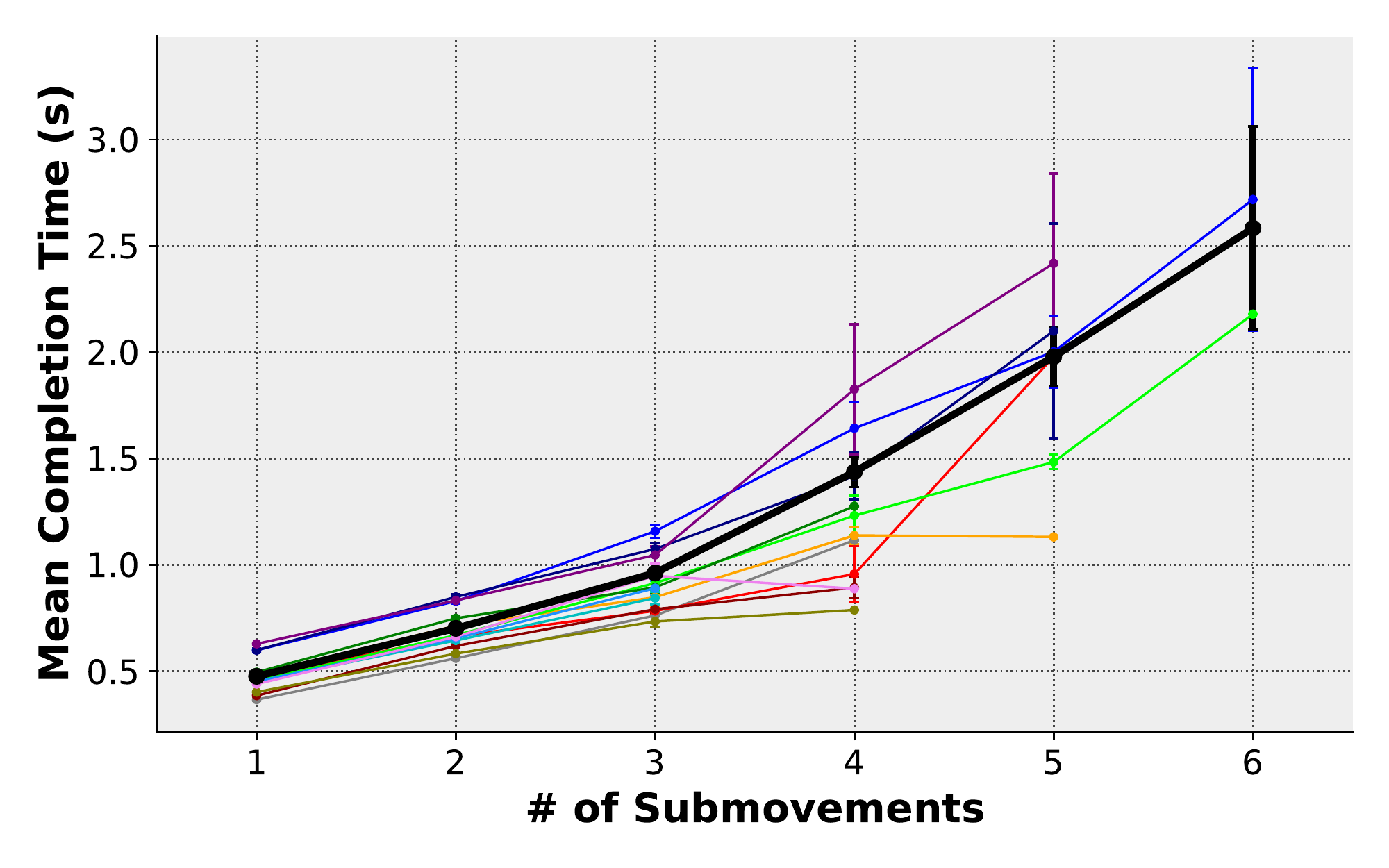}
    \vspace{-4mm}
    \caption{(Left) average velocity of the first submovement versus mouse sensitivity, demonstrating that though high sensitivity did degrade overall task time, it did not as substantially effect player movement velocity. (Center) positive correlation between sensitivity and submovement count. (Right) overall positive correlation between submovement count and average task completion time. Note that since subjects were less likely to make higher numbers of submovements (standard error) error bars grow substantially for large submovement counts.}
    \vspace{-4mm}
    \label{fig:SubmovementSpeedCount}
\end{figure}

By partitioning player motion into submovements, and only analyzing the first submovement of the series (typically  the largest motion) we can better separate slower performance due to long initialization or verification windows from slower average player motion.
This analysis demonstrates that for the lower, and in some cases the highest, tested sensitivities, players rotated their view more slowly in game.
However, the degradation of movement speed at high sensitivities is far less than that of the high sensitivities in Fig. \ref{fig:MeanTaskTimeThroughput}. Instead, the average number of submovements monotonically increases with sensitivity (Fig. \ref{fig:SubmovementSpeedCount} center plot). 
This indicates that targeting motion precision degrades with mouse sensitivity, requiring more corrective submovements before successfully completing a targeting action.
This increase in submovement count drives an increase in mean task completion time in turn, as demonstrated in the rightmost plot of Fig. \ref{fig:SubmovementSpeedCount}.
Therefore, the optimal region in task completion time, throughput, and average view rotation speed in the plots in Fig. \ref{fig:MeanTaskTimeThroughput} are not solely the result of increased player motion speed, but rather the trade-off of movement speed impacts at low mouse sensitivity, and increased time for small, corrective movements at high sensitivities.


\section{Discussion}
\label{sec:Discussion}
Generally speaking, the range of optimal suggested mouse sensitivity for FPS aiming tasks in this work conforms to the range suggested by previous work.
The near optimal 4x range of sensitivities arrived at in our study matches the multiplicative range of optimality arrived at in prior art.
With this in mind we revisit our 3 experimental design questions.

\subsection{Experimental Questions}
Our experiment design asked three fundamental questions about competitive gamers' mouse sensitivity settings:
\begin{itemize}
    \item Does an optimal sensitivity range exist for competitive FPS targeting tasks?
    \item Are competitive FPS gamers preferred sensitivities well aligned with their optimal performance?
    \item Does (long term) muscle memory play a crucial role in aiming performance? If so, does playing with varied sensitivity significantly degrade performance?
\end{itemize}

Based on our measured and modeled results we answer these questions as follows.

\paragraph{Optimal Sensitivity Range}
Our competitive FPS gamers did demonstrate a preference for an overall optimal range of mouse sensitivity, a 4x range from 0.45-1.8$^\circ$/mm or 20-80 cm/360$^\circ$.
This 4x range of near optimal sensitivity mirrors that of previous work in CD gain based pointer-based literature, suggesting a possible common basis for this range of near optimal performance.

\paragraph{Individually Optimal Sensitivity}
The gamers we studied did not demonstrate significantly \emph{better} choice of preferred sensitivity than generally optimal ones; however, all but two did use preferred sensitivities within the globally optimal region.
The two users who selected sensitivities outside of (higher than) the optimal range were among the poorest performers, but still demonstrated less degradation in performance near these choices than other users.

\paragraph{Long-term Effects of Sensitivity Changes}
\begin{figure}
    \centering
    \includegraphics[width=0.54\textwidth]{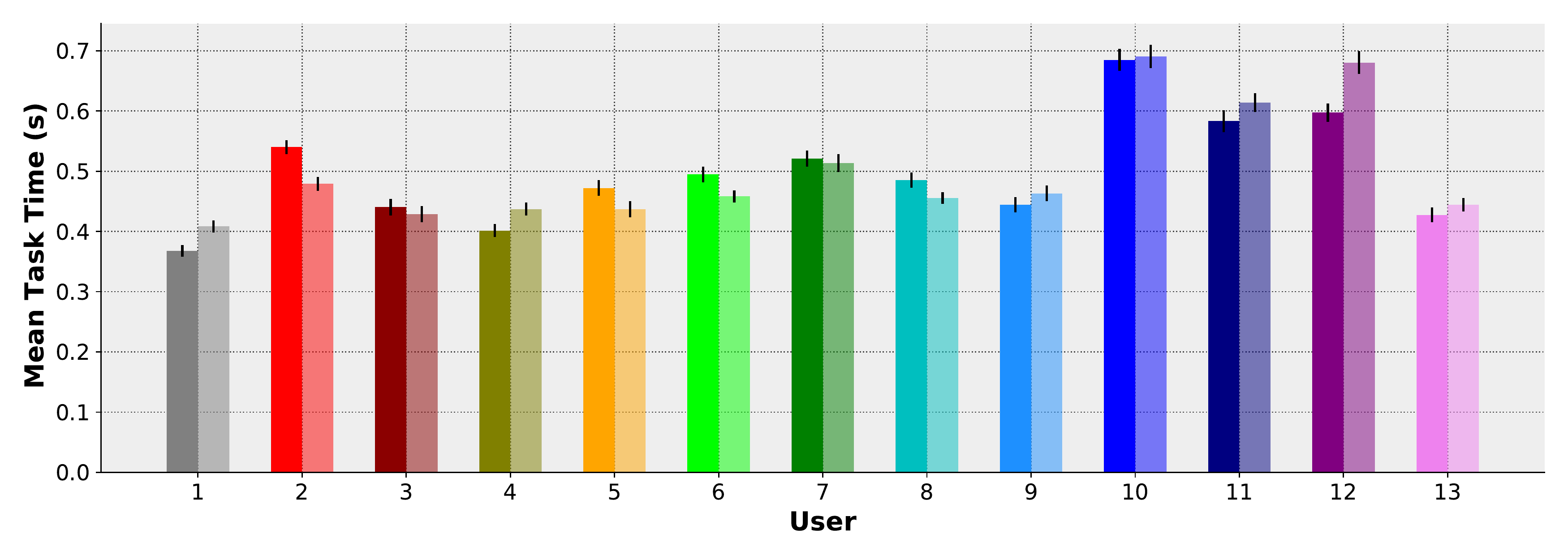}
    \includegraphics[width=0.45\textwidth]{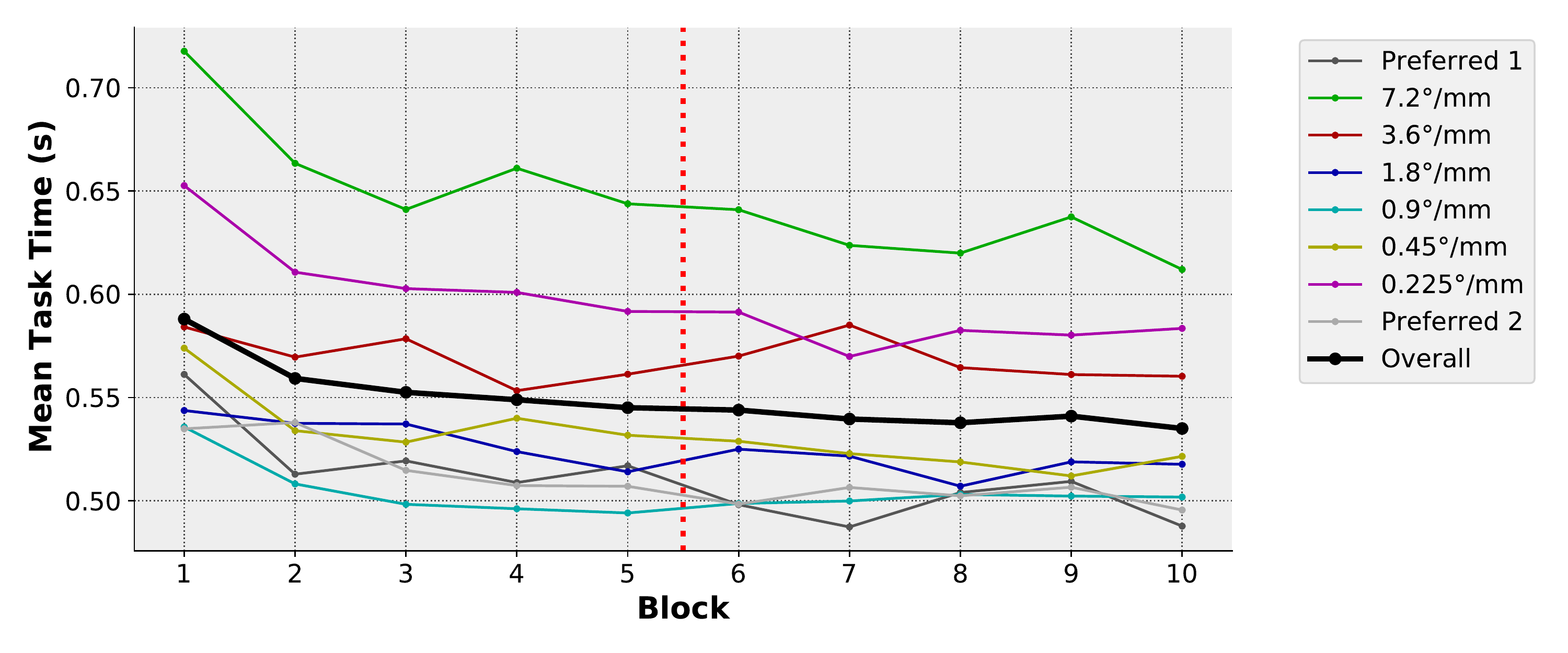}
    \vspace{-4mm}
    \caption{(Left) comparison of average task completion time between the first (left bar) and final (right bar) session, both at the preferred sensitivity, by user. Error bars are standard error. Some users show statistically significant degradation of performance (increase from left to right bar), while other show no significant change, or even small significant improvement. (Right) per-session average completion time over blocks of 50 trials demonstrating the short-term training effect. The dashed vertical red line represents the threshold for discarding trials due to short-term training effects.}
    \label{fig:TrainingEffects}
    \vspace{-4mm}
\end{figure}

While short-term muscle memory was demonstrated by a strong within-session training effect as demonstrated in Fig. \ref{fig:TrainingEffects} (controlled for by only analyzing the last 50\% of trials from any given condition), longer-term (i.e., session-to-session) training effects were not as pronounced in our data.
Overall our subjects did not demonstrate a significant degradation of performance between their first (preferred) sensitivity session, and their final sensitivity session.
Individually, about half of subjects demonstrate degradation of performance, an increase in mean task time, from the first trial to the last, but nearly as many showed \emph{improvement} between these same sessions.
Generally, this degradation/improvement was small in magnitude (<40\% of the standard deviation of completion time, and < 15\% of average completion time) and statistically insignificant.
Interestingly the subjects with the highest and lowest average completion times, or the least and most skilled users, tended to experience this degradation.
This may imply that the most skilled subjects did experience a performance detriment from changing sensitivities, while slightly less skilled players benefited more generally from time spent in the experimentation platform.

\subsection{Suggestions for Mouse Sensitivity Selection}
\vspace{-1mm}
\label{sec:RecommendedSensitivity}
In regard to selecting a mouse sensitivity for FPS gaming, our suggestions largely fall in line with historical work in sensitivity recommendations based on spatial task difficulty.
This entails identifying the common ranges of (angular) target size and distance (and thus their corresponding IDs) and selecting a mouse sensitivity that optimizes the trade-off of speed of motion to number of submovements (i.e. chance of over/undershoot) for this sensitivity.
From the gamer's perspective this means balancing mouse sensitivity high enough that physical motion doesn't limit in-game movement speed, but low enough that over/undershooting targets isn't too common of an  occurrence.
For our study we attempted to cover a wide but representative range of IDs common in FPS gaming, but in reality individual gamers may choose to further taper their sensitivity to their role (i.e. tasks) in-game.

As an example let us consider two roles common within a single competitive FPS game: a player who interacts with a target at a short distance and a player who interacts with the same target at a much greater distance, with the target distributed over a similar range of view eccentricity in both cases.
Since the target remains the same size but is presented at different depths, the first player experiences a larger target size relative to the aim displacement required to hit it, while the second player must select a smaller target at a similar aim displacement.
This means the first player experiences tasks with lower ID, while the second player experiences tasks with consistently higher ID.

\begin{figure}
    \centering
    \includegraphics[width=0.95\textwidth]{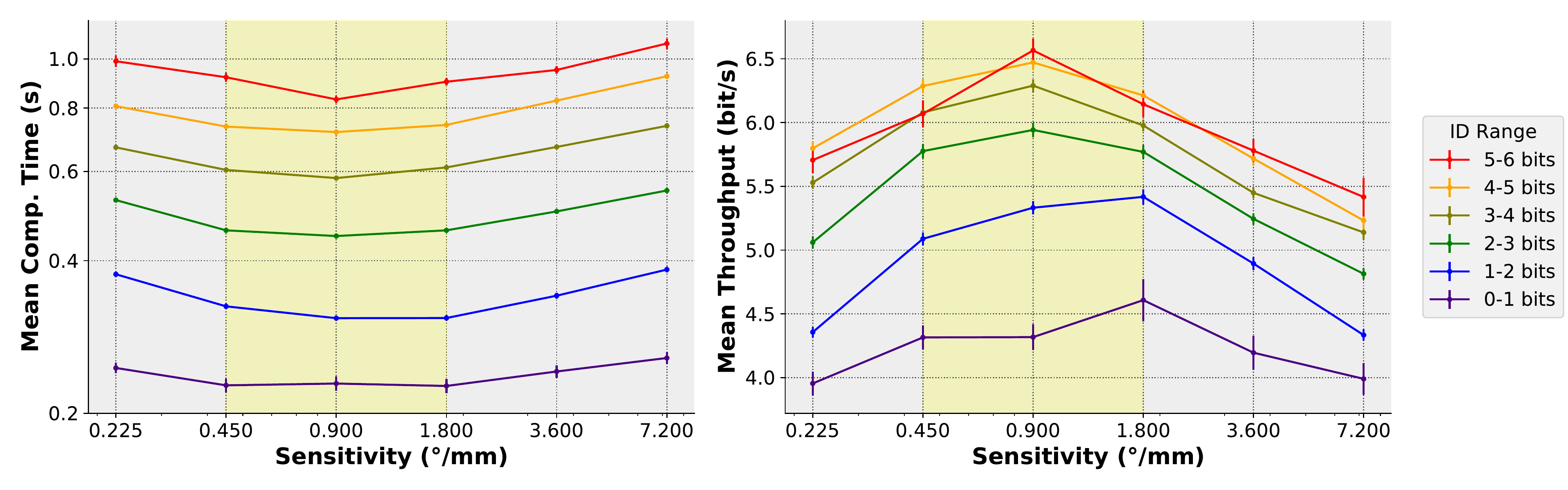}
    \vspace{-4mm}
    \caption{Average task completion time (left) and throughput (right) for all participants plotted by various ID ranges (colors). A trend towards higher sensitivity at lower spatial task difficulty is observable. The mean completion time is plotted with log X and Y axes. Throughput is plotted with a log X axis. Error bars represent standard error.}
    \label{fig:TimeThroughputByID}
    \vspace{-6mm}
\end{figure}

Per traditional user interface recommendations, the player interacting with low ID (nearby) targets should select a higher mouse sensitivity, allowing them to rotate the view more quickly towards these targets, and make rapid changes in view direction to track nearby targets.
This player's risk of under/overshooting is greatly reduced by the increased target size they experience.
The player experiencing higher ID (more distant) targets stands to benefit from lower mouse sensitivity, as since the need to make sudden, dramatic changes in view direction is lower this player benefits more from the increased precision of lower sensitivity.
We demonstrate this claim by plotting our task completion time and throughput data broken out by range of IDs in Fig. \ref{fig:TimeThroughputByID}.
The results demonstrate subjects performed better at lower sensitivities for high ID tasks, and higher sensitivities for low ID tasks, in accordance with our suggestion and prior art. 

Notable exceptions to this principle arise when display-space boundaries (such as display borders) are used to "stop" mouse travel at some extent.
This example non-linearity is utilized in many OS-level interfaces (e.g., the start menu button) which allows easier selection of a small but distant (high ID) target by limiting pointer travel in the corner of the display.
Such a limitation applies in FPS aiming as the view elevation (vertical-displacement) angle is often limited to $\pm90^\circ$.
This implies targets located directly above or below the player view position may be substantially easier to target even if they have a high ID.


\subsection{Suggestions for Future Work}
\vspace{-1mm}
The range of target widths and distances studied in this work represents a reasonable sampling of ID, but a subset of targeting tasks presented in real FPS games. Studying the impacts of sensitivity in large angular rotations (i.e. 180$^\circ$ turns in game) may present an added benefit to lower mouse sensitivities not represented in this work. This may be more relevant to some game conditions than others.

In addition, many games implement separate sensitivities for "scoped" or "zoomed" modes that modify aiming behavior, particularly when trying to hit small targets at reasonable distances. 
Our study did not include this ability, but a future study could include this dynamic and study how the ability to select between two different target sizes/difficulties changes the optimal behavior, at the cost of field of view while zoomed in.

Typically gamers learn the limits of their mouse pad or surface and tend to lift (or clutch) the mouse prior to these limits, creating additional non-linearity in task completion time data.
The mouse motion studied in this work occurs over a range of  ~100 mm (the maximum target distance at the lowest sensitivity corresponds to 111.8 mm of mouse travel) to discourage subjects from lifting their mouse during trial data collection.
However, gaming mouse pads can be as large as 400-500 mm~\cite{prosettings-mousepads}.
Future work exploring larger target distances (i.e. user motion $>$500 mm) would encourage more lifting actions, improving understanding of how their speed and duration is related to target task difficulty and sensitivity.

\section{Conclusions}
We conduct the first performance-oriented study of mouse sensitivity in FPS targeting tasks and report a range of generally optimal sensitivities from 0.45-1.8$^\circ/$mm.
This agrees well with the 4x range of CD gain recommended by prior art in pointer-based mouse sensitivity literature, as well as the preferred sensitivities selected by our experienced FPS gamer subjects.
We show that low sensitivities reduce in-game view rotation speed while high sensitivities result in reduced motion precision (increased corrective motions), both increasing overall task completion time.
In addition, we demonstrate that low ID tasks benefit from higher sensitivities, while more spatially difficult (high ID) tasks benefit from lower sensitivities, again conforming to traditional pointer-based results.
Finally we did not observe long-term "hard wired" muscle memory of mouse sensitivity for all players, and we encourage FPS gamers to experiment with changing mouse sensitivity to continue to optimize their settings to their role and individual play style.

\begin{acks}
The authors acknowledge Ben Watson, Byungjoo Lee, and Rachel Brown for comments on the ideas behind this work.
Thanks to Justin Thomas for administering the user study.
\end{acks}

\bibliographystyle{ACM-Reference-Format}
\bibliography{main}

\end{document}